\documentclass[journal]{IEEEtran}
\usepackage{graphicx}
\usepackage{subfigure}
\usepackage{float}
\usepackage{amsmath}
\usepackage{amsfonts,amssymb}
\usepackage{dsfont}
\usepackage{algpseudocode}
\usepackage{makecell}
\usepackage{cite}
\usepackage{multirow}
\usepackage{multicol}
\usepackage{booktabs}
\usepackage[ruled,vlined]{algorithm2e}
\usepackage{xcolor,soul}
\usepackage{comment}

\DeclareFontFamily{U}{mathx}{}
\DeclareFontShape{U}{mathx}{m}{n}{<-> mathx10}{}
\DeclareSymbolFont{mathx}{U}{mathx}{m}{n}   
\DeclareMathAccent{\widetilde}{0}{mathx}{"72}

\usepackage{fancyhdr,lipsum}

\fancypagestyle{mahmood}{%
   \fancyhf{} 
   
   \fancyhead[C]{You can use this material personally. Reprinting or republishing this material for the purpose of advertising or promotion, creating new collective works, reselling or redistributing to servers or lists, or using any copyrighted component in other works must adhere to IEEE policy.}
}%

\makeatletter
\let\ps@IEEEtitlepagestyle\ps@mahmood
\makeatother

\begin{document}
\title{Towards a Dynamic Future with Adaptable Computing and Network Convergence (ACNC)}

\author{Masoud Shokrnezhad\textsuperscript{1, 2}, Hao Yu\textsuperscript{1}, Tarik Taleb\textsuperscript{3}, Richard Li\textsuperscript{4}, Kyunghan Lee\textsuperscript{5}, \\ Jaeseung Song\textsuperscript{6}, and Cedric Westphal\textsuperscript{7}
\\
\textsuperscript{1} \textit{ICTFicial, Finland; \{firstname.lastname\}@ictficial.com} \\
\textsuperscript{2} \textit{Oulu University, Oulu, Finland; masoud.shokrnezhad@oulu.fi} \\
\textsuperscript{3} \textit{Ruhr University Bochum, Bochum, Germany; tarik.taleb@rub.de} \\
\textsuperscript{4} \textit{Southeast University, Nanjing, China; richard.li@seu.edu.cn} \\
\textsuperscript{5} \textit{Seoul National University, Seoul, Korea; kyunghanlee@snu.ac.kr} \\
\textsuperscript{6} \textit{Sejong University, Seoul, Korea; jssong@sejong.ac.kr} \\
\textsuperscript{7} \textit{Futurewei Technologies, USA; cedric.westphal@futurewei.com}

}




\maketitle

\begin{abstract}
In the context of advancing 6G, a substantial paradigm shift is anticipated, highlighting comprehensive everything-to-everything interactions characterized by numerous connections and stringent adherence to Quality of Service/Experience (QoS/E) prerequisites. The imminent challenge stems from resource scarcity, prompting a deliberate transition to Computing-Network Convergence (CNC) as an auspicious approach for joint resource orchestration. While CNC-based mechanisms have garnered attention, their effectiveness in realizing future services, particularly in use cases like the Metaverse, may encounter limitations due to the continually changing nature of users, services, and resources. Hence, this paper presents the concept of Adaptable CNC (ACNC) as an autonomous Machine Learning (ML)-aided framework crafted for the joint orchestration of computing and network resources, catering to dynamic and voluminous user requests with stringent requirements. ACNC encompasses two primary functionalities: state recognition and context detection. Given the intricate nature of the user-service-resource space, the paper employs dimension reduction to generate live, holistic, abstract system states in a hierarchical structure. To address the challenges posed by dynamic changes, Continual Learning (CL) is employed, classifying the system state into contexts controlled by dedicated ML agents, enabling them to operate efficiently. These two functionalities are intricately linked within a closed loop overseen by the End-to-End (E2E) orchestrator to allocate resources. The paper introduces the components of ACNC, proposes a Metaverse scenario to exemplify ACNC's role in resource provisioning with Segment Routing v6 (SRv6), outlines ACNC's workflow, details a numerical analysis for efficiency assessment, and concludes with discussions on relevant challenges and potential avenues for future research.
\end{abstract}
\begin{IEEEkeywords}
6G, Deterministic Networking, Computing-Network Convergence (CNC), Joint Resource Allocation, Metaverse, Autonomous Orchestration, Dimension Reduction, Context Detection, Continual Learning, Machine Learning, and Segment Routing v6 (SRv6).  
\end{IEEEkeywords}


\IEEEpeerreviewmaketitle

\section{Introduction}\label{s_intro}
In the foreseeable future for the 6th generation of communication systems (6G),  interrelations are anticipated to enlarge beyond purely human-to-human connections to include everything-to-everything interactions. The Metaverse, a virtual realm of boundless possibilities, is poised for an era of rapid and explosive expansion. In this digital universe, a diverse multitude of both physical and digital objects will incessantly engage, traverse, and coexist within a seamlessly interconnected global network. This dynamic and immersive environment is expected to be further amplified by the persistent growth of connected objects, giving rise to a prolific exchange of data and generating substantial volumes of upstream traffic. In essence, the presence of billions of entities operating at exceptionally high densities, along with the generation of zettabytes of global data, is expected to be the norm of 6G \cite{tang_computing_2021}. This explosive growth in demand must be acknowledged, particularly as a substantial portion of this traffic will need to be transmitted to computing nodes for processing before being returned to users.

An additional aspect of the swift progress of technology pertains to the wide spectrum of promising and innovative services expected to materialize in the near future. One notable example is the concept of \textit{Metaverse as a Service (MaaS)}, empowering users to create, host, and manage personalized virtual environments across various domains like healthcare, finance, entertainment, and education. When combined with other capabilities, such as \textit{Spatial Computing}, facilitating the convergence of physical and digital realms through \textit{Virtual}, \textit{Augmented}, and \textit{Extended Realities} (\textit{VR/AR/XR}) \cite{10293194} to create a \textit{Digital Twin of a Person (DToP)}, these combined capabilities yield hybrid services replicating an individual's distinctive attributes and enabling their coexistence in multiple locations, bridging both digital and physical domains simultaneously. It is essential to emphasize that these services necessitate strict adherence to Quality of Service/Experience (QoS/E) standards. The realization of these services hinges on several Key Performance Indicators (KPIs), which include achieving millisecond-level End-to-End (E2E) latency, an ultra-high reliability rate, and a peak data rate in the terabit range.

\begin{figure*}[!t]
\centerline{\includegraphics[width=7.1in]{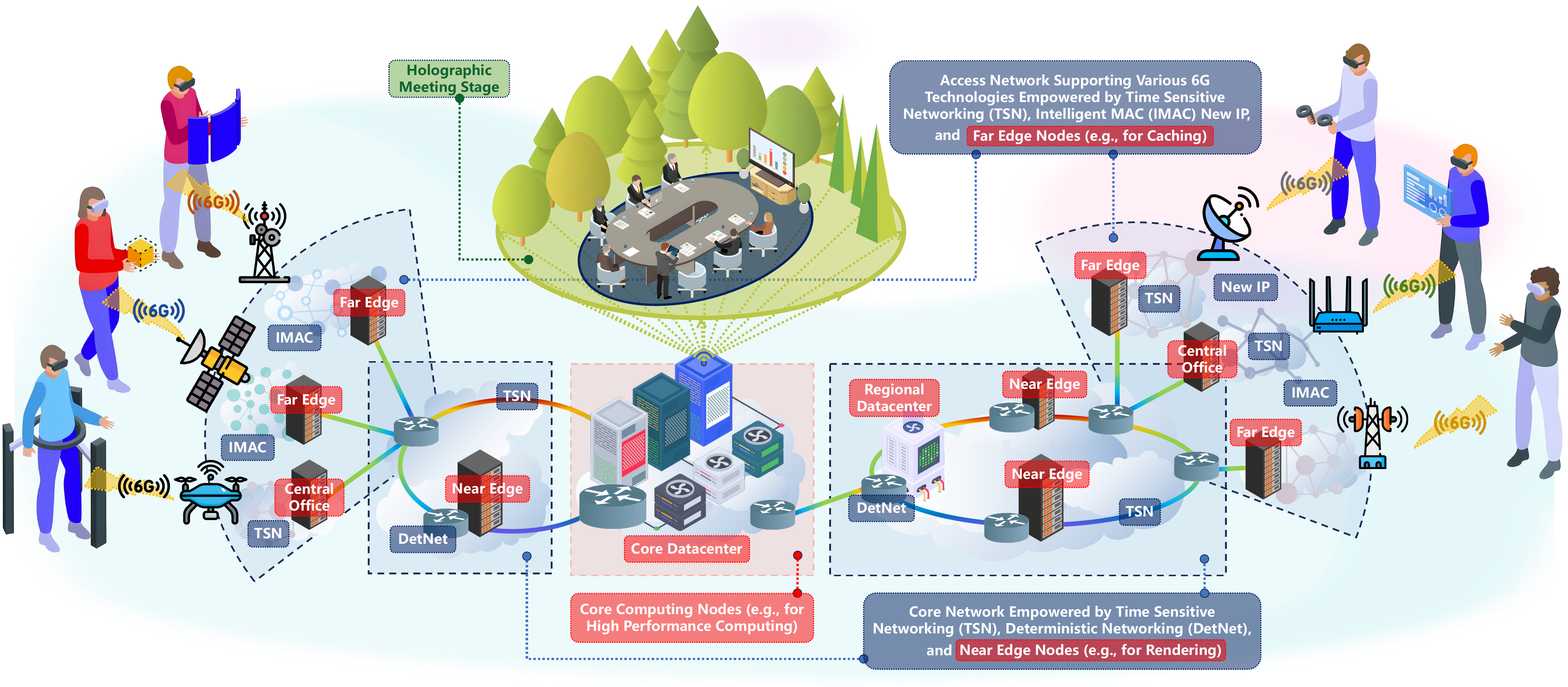}}
\caption{A lakeside holographic meeting room in the Metaverse enabled by a cloud-network integrated infrastructure powered by technologies including deterministic networking, time-sensitive networking, and intelligent medium access control, with users connected via 6G connections \cite{shokrnezhad2023TMCCLMA}.}
\label{fig1}
\end{figure*}

The primary challenge in realizing the envisioned future lies in addressing the substantial demands on computing and network resources, coupled with stringent QoS/E requirements, stemming from inherent resource scarcity. A promising strategy to effectively address this challenge involves shifting from the prevailing paradigm of \textit{"user $\leftrightarrow$ network $\leftrightarrow$ computing"} to a more integrated approach denoted as Computing-Network Convergence (CNC), symbolized by \textit{"user $\leftrightarrow$ $<$network $\oplus$ computing$>$"}. This paradigm shift has garnered significant attention in recent years. Król \textit{et al.} \cite{krol_compute_2019} explored this concept through Compute-First Networking (CFN), applying Information-Centric Networking (ICN) principles to automatically perform traffic routing on network devices based on data object names. Two complementary approaches, Compute-Aware Networking (CAN) \cite{han_utility-optimized_2021} and Computing Power Networking (CPN) \cite{tang_computing_2021}, empower network devices to include computing nodes' capabilities and load statuses in their decisions. Zhang \textit{et al.} \cite{zhang_research_2022} proposed a comprehensive CNC architecture integrating resource allocation for both computing and network domains. Additionally, Albalawi \textit{et al.}~\cite{albalawi2019inca} investigated a concept similar to CNC, termed Integrated Network and Compute Allocation (INCA).

The effectiveness of the mentioned joint-orchestration strategies may face limitations, particularly in use cases like the Metaverse. An extra complication is the dynamic nature of user requirements, seamless transitions between services in virtual realms, and frequent traversal between virtual and physical environments. In such a dynamic ecosystem, the live and scalable distribution of complex system states to network devices, coupled with resource allocation decision-making, becomes crucial. Existing approaches are unsuitable for this task due to significant convergence times, potentially compromising stringent QoS/E requirements amidst numerous dynamic users and services \cite{tang_computing_2021, sun_computing_2022, 9766416}. This paper addresses a gap in the existing literature by proposing a Machine Learning (ML)-based solution for the joint orchestration problem, considering the dynamic nature of user behavior, service characteristics, and infrastructure resources. Our proposed solution, Adaptable CNC (ACNC), seamlessly integrates ML-based state recognition and context detection across resources, as well as domain and E2E orchestrators. Utilizing hierarchical state recognition, ML-based reduction techniques efficiently handle the complexity of extracting system states from the vast user-service-infrastructure space for resource allocation. Through context detection, dynamicity is handled by categorizing reduced yet ubiquitous system states into predefined contexts, enabling resource allocation agents dedicated to each context to efficiently converge to optimal solutions over similar states while maintaining divergence from distinct states.

The subsequent sections are organized as follows: Section \ref{s_pot_sul} delves into the problem at hand, elucidating the dynamic nature of future use cases and the key role of Continual Learning (CL) in the ACNC framework. Section \ref{s_cnc} provides a comprehensive description of ACNC, including its architecture, components, workflow, and a scenario within the Metaverse. Performance assessment is covered in Section \ref{s_eva}, followed by a discussion on potential challenges and future research directions in Section \ref{s_chl}. The paper concludes with final remarks in Section \ref{s_con}.

\section{Fundamentals}\label{s_pot_sul}

\subsection{The Interwoven Cloud-Network Harmony}\label{ss_6g}
In the landscape of future 6G systems, a vision emerges of integrated, decentralized computing resources spanning in-network, edge, regional, and central nodes, interconnected by robust networking technologies. The primary challenge for facilitating extensive-scale services with copious data throughput, ultra-broadband connectivity, exceptional reliability, and minimal latency lies in the precise orchestration of these resources. Illustrated in Fig~\ref{fig1} is a Metaverse scenario where users engage in a holographic presence service, participating in a conference room virtually located on a live lake shore. To realize this experience, a sequence of steps is undertaken. Initially, necessary service instances, including rendering, motion tracking, stereoscopic 3D display, and audio spatialization, are instantiated on available computing nodes. Subsequently, a comprehensive dataset, encompassing video, audio, and motion data, is meticulously captured from each user and injected into the service instances following a predefined order specified in the service's function chaining map. Finally, the resulting rendered content is transmitted to all users for display on their respective devices, such as headsets.

Traditional resource orchestration methodologies are insufficient for organizing such integrated infrastructures to execute such complex scenarios, as they typically operate within individual domains, neglecting interdependencies among diverse domains and resources. In contrast, CNC-based approaches adopt a holistic perspective, considering the web of interactions for optimal E2E orchestration and efficient resource utilization. In the holographic conference scenario, CNC-based orchestration can integrate network link status, strategically placing service instances on computing nodes to leverage shared links and minimize network utilization for high-bandwidth requests. Additionally, they can mitigate ripple effects from failures or limited functionality within one domain on the other. For instance, in the event of congestion in allocated network links, CNC-based approaches can reposition service instances to open new, efficient networking opportunities, maintaining strict QoS/E requirements. This adaptive approach contrasts with conventional methods that attempt to navigate congested links and seek alternative but merely feasible paths to the same instance locations.

\subsection{The Horizon of Ever-Changing Services}\label{ss_dynamic}
Despite the commendable performance of joint-orchestration strategies, they prove inadequate in use cases like the Metaverse, where the system experiences constant and multidimensional fluctuations. In the virtual conference scenario, users may physically navigate, adjusting lighting or modifying the scene on their virtual 3D monitors, requiring rapid adaptations in data collection and QoS/E requirements. In another example, introducing users with diverse language preferences may necessitate distinct service instances for audio sampling, translation, and reconstruction. Moreover, attendees may transition between virtual settings, such as altering the visual platform from the lake shore to the assembly line of their factory so that they may conduct an in-depth analysis of processes, requiring changes in loaded virtual environments. Dynamism is also introduced by the underlying infrastructure network devices and computing nodes, dispersed and operating under distinct administrations with fluid characteristics. To effectively manage these ever-changing situations, CNC-based approaches must swiftly reorganize changes, often within microseconds, ensuring seamless system operation.

\subsection{The Wisdom of Continuous Conetx-Awareness}\label{ss_lifelong}
Efficient resource allocation in dynamic, complex systems requires adaptive strategies that can anticipate changes and respond appropriately in real-time. Traditional ML approaches frequently face challenges in such environments due to their inability to effectively manage non-stationary data. The acquisition of new information can undermine previously learned knowledge, leading to decreased performance on earlier tasks, a phenomenon known as \textit{catastrophic forgetting}. CL presents a promising solution by enabling intelligent agents to continuously acquire new knowledge while retaining previously learned information, ensuring accurate decision-making across diverse system states \cite{verwimp2023continual}. To facilitate this, \textit{contexts} are introduced in CL, which partition various system states into distinct categories, each representing a specific configuration or operational condition. By organizing system states into finite contexts with dedicated training processes, agents can specialize in particular scenarios without interference, thereby mitigating catastrophic forgetting and enhancing adaptability and responsiveness. Furthermore, the modularity of this approach allows agents to be trained in parallel across different contexts, utilizing parallel computing resources to reduce training time. In this paper, we propose that the current context can be defined by the existing service instances on computing nodes. The introduction of new instances or the migration or removal of existing ones facilitates the transition to a new context. For example, $\mathcal{C}_1: \{\mathcal{N}_1 \gets \mathcal{I}_1, \mathcal{N}_2 \gets \mathcal{I}_2\}$ and $\mathcal{C}_2: \big\{\mathcal{N}_1 \gets \{\mathcal{I}_3, \mathcal{I}_4\}, \mathcal{N}_2 \gets \mathcal{I}_1\big\}$, where $\mathcal{C}_i$ denotes context \(i\), $\mathcal{N}_j$ represents computing node \(j\), and $\mathcal{I}_k$ indicates service instance \(k\).

\begin{figure*}[!t]
\centering
\includegraphics[width=1\textwidth]{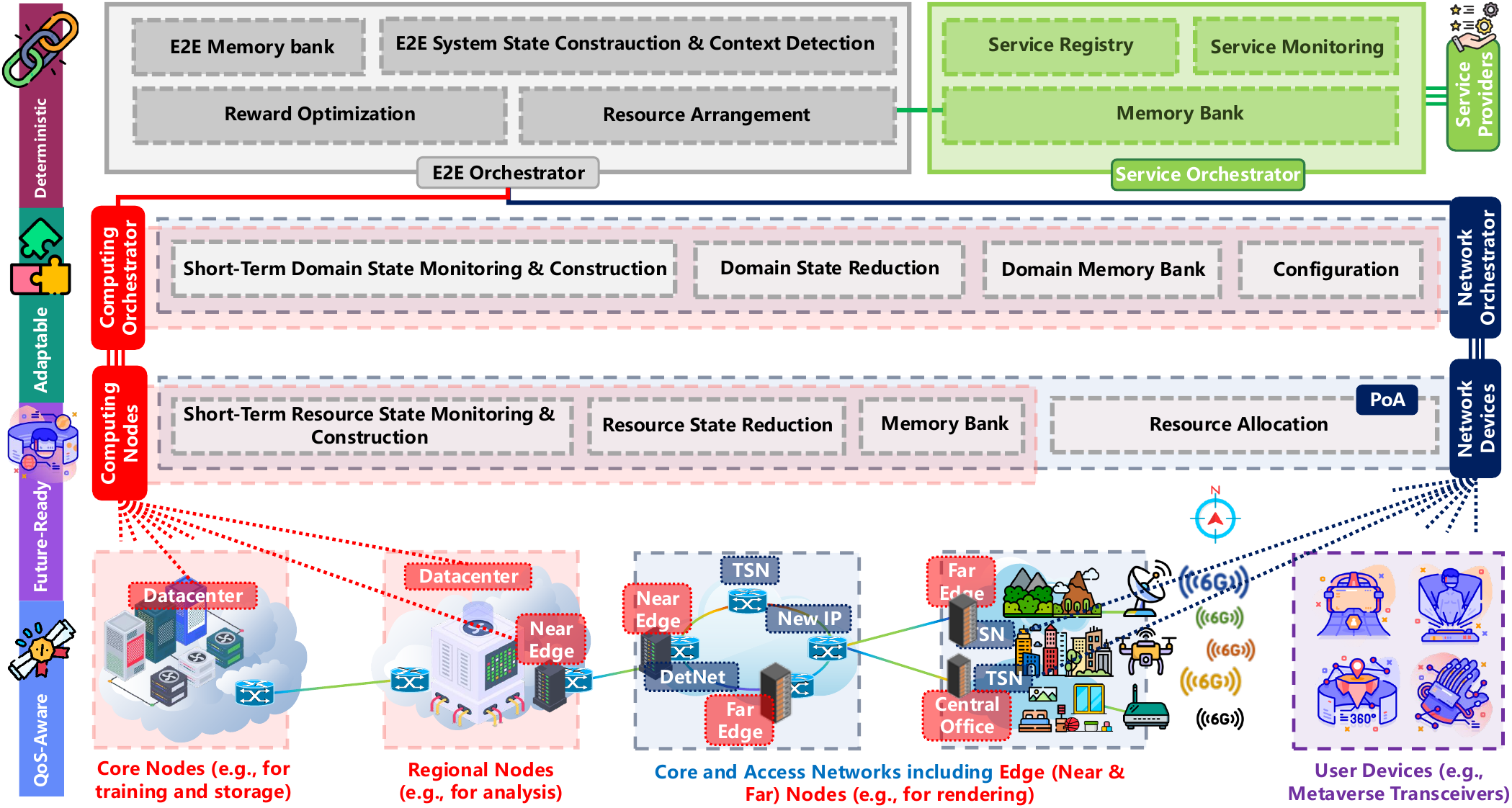}
\caption{
The Architecture of Adaptable Computing-Network Convergence (ACNC).
}
\label{fig2}
\end{figure*}

\section{Adaptable Computing-Network Convergence}\label{s_cnc}

\subsection{The Parties Involved}
This paper involves three key parties. The first party, termed \textit{the infrastructure}, encompasses the integrated computing and network resources of future 6G systems, along with their orchestration components. We assume the infrastructure relies on established protocols and data formats, ensures that data exchanged between different systems are secured and accurately interpreted. The second party is \textit{the service provider}, responsible for registering services within the system. These services, illustrated by scenarios like the holographic meeting, may involve single or multiple instances that communicate according to a predefined sequence and data format outlined in the service's function chaining map specified during registration. The ultimate participant in the system is \textit{the user}, responsible for submitting requests to access registered services. Each request requires an E2E connection, establishing a link between the user and the relevant service instance(s). This connection manages the transmission of data from diverse sources (e.g., microphones, cameras, positioning sensors) to the instances and delivers processed data, such as the live-rendered holographic meeting scene, back to the user.

\subsection{The Architecture of Adaptable Converged Orchestration}\label{ss_arch}
With the consideration of the involved parties, the core challenge involves the placement of registered service instances onto computing nodes while simultaneously establishing E2E paths that facilitate request connectivity and traversing the required service instances, adhering to stringent QoS/E criteria. The overarching objective is to address this challenge in order to accommodate the systems's dynamic nature. To meet this, we introduce a framework named ACNC, representing the integration of CL principles with the CNC concept, with two primary principles: 1) optimizing infrastructure capacity utilization through joint orchestration to maximize the number of supported users, and 2) simplifying the complexity by defining discrete contexts, thereby enabling proactive readiness to tackle context-specific fluctuations. The proposed framework consists of various layers and components is depicted in Fig. \ref{fig2} and defined in what follows.

\vspace{10px}
\subsubsection{Resources}\label{sss_res} \hfill \vspace{3px} \\  \indent 
In the ACNC framework, each network device's primary function is the creation of short-term states. This involves meticulous processing of traffic across all ports, where metrics like traffic characteristics, request age of information, and QoS/E requirements are measured and collected for all requests. The short-term resource state at time slot $t$ is denoted as $\mathcal{\varphi}^t_R$, with a spatial dimensionality represented as $|\mathcal{\varphi}^t_R|: \mathcal{U} \times \mathcal{S} \times \mathcal{R} \times \mathcal{P} \times \mathcal{M}_R$\footnote{Let $\mathcal{U}$ denotes the number of users, $\mathcal{S}$ signifies the number of active services, $\mathcal{R}$ represents the quantity of requests per user and service, $\mathcal{P}$ indicates the number of ports, and $\mathcal{M}_R$ stands for the quantity of metrics measured per request.}. To comprehend causal relationships between resource allocation decisions and state transitions, short-term resource states for time slots $[t-\mathcal{T}_R, t]$ are aggregated to form a long-term resource state at time slot $t$ ($\widehat{\mathcal{\varphi}}^t_R$). Given the complexity and multidimensionality of this state ($|\widehat{\mathcal{\varphi}}^t_R|: |\mathcal{\varphi}^t_R| \times \mathcal{T}_R$), we utilize ML techniques, including Uniform Manifold Approximation and Projection (UMAP), a method known for its ability to handle large datasets, to reduce dimensions. The resulting reduced resource state is denoted as $\widetilde{\mathcal{\varphi}}^t_R$, with size $\mathcal{S} \times \mathcal{P} \times \widetilde{\mathcal{M}}_R$, where $\widetilde{\mathcal{M}}_R$ represents the size of the set representing each service over each port that can accommodate from one indicator to all \(\mathcal{M}_R\) metrics. A smaller set of indicators yields more concise reduced states, simplifying state exchanges but diminishing precision in state representations. After reduction, the device stores the reduced resource state in the memory bank and then forwards it to the network orchestrator. Given the reduced resource state size, the required time and resources for this exchange are predictable. The same operations are replicated on computing nodes, but they construct states for each service instance implemented on their infrastructure.

\begin{figure*}[!t]
\centering
\includegraphics[width=1\textwidth]{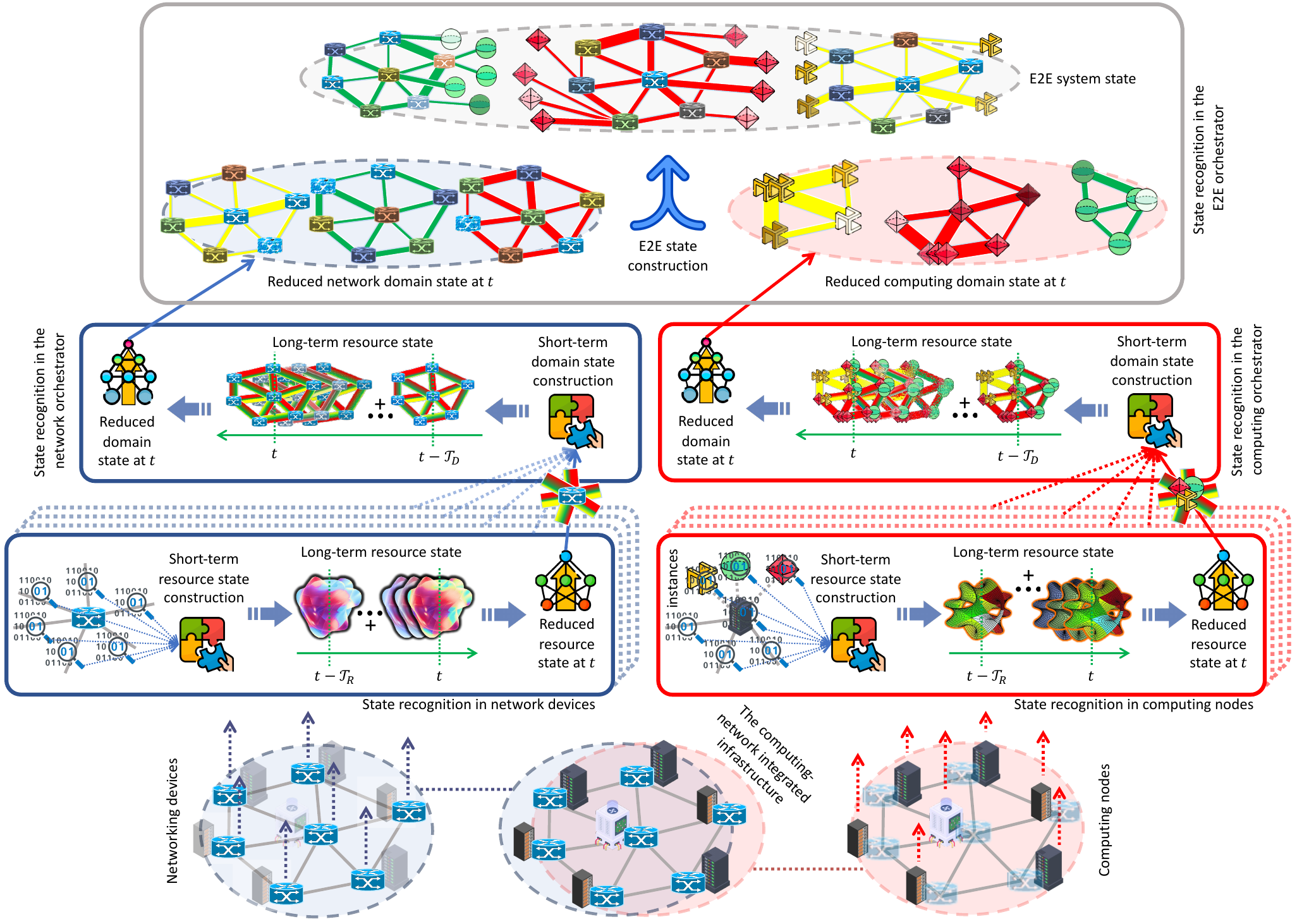}
\caption{The process of state construction and reduction within the ACNC framework.}
\label{fig3}
\end{figure*}
For each edge network device or Point of Arrival (PoA), there is an additional task - the selection of service instances for incoming requests and determining their respective network paths. This task is executed in a context-aware manner, where each PoA is equipped with an agent dedicated to a unique context. These contexts are structured to enable a singular agent to proficiently oversee the dynamics of the system states associated with that context, leading to a notable reduction in overall complexity to a more manageable scale. The agent for each context maintains two action sets: 1) the available service instances within that context, and 2) the available network paths from the PoA to these service instances. When the context changes or a significant alteration occurs in the system state, the PoA receives relevant information about the active context and current system state in graph form (as detailed in Section \ref{sss_e2e}) from the network orchestrator. Subsequently, the PoA activates the agent associated with the active context, assigning service instances and network paths to requests according to their chaining map within the specified time slot. We employ a Reinforcement Learning (RL) framework to implement this agent, wherein the system state and request information serve as inputs. The agent takes actions (instances and paths) to maximize a cumulative reward signal over time (see Section \ref{sss_e2e}). To process the input effectively, the agent is implemented using Graph Neural Networks (GNNs) \cite{10161704}. The GNN aggregates information from neighboring nodes within the system state graph, allowing the agent to incorporate both local context and global state conditions into its decision-making process.

\vspace{10px}
\subsubsection{Domain Orchestrators} \hfill \vspace{3px} \\  \indent 
The orchestration layer consists of two key components: the network and computing orchestrators. Both orchestrators aggregate reduced resource states from members in their respective domains, establishing a comprehensive short-term domain state for each time slot ($\varphi^t_D$). The network orchestrator constructs a graphical representation for each service, reflecting the physical network's port-level structure. Simultaneously, the computing orchestrator formulates an overlay map for each service based on its service instance chain. Moreover, the orchestrators monitor resource availability by collecting metrics such as operational uptime, failure events, resource utilization rates, power supply statuses, and environmental conditions, adding them to $\varphi^t_D$, with the number of metrics denoted by $\mathcal{M}_D$. Now, aggregating short-term domain states for $[t-\mathcal{T}_D, t]$ time slots, both orchestrators employ dimension reduction ML models to create reduced domain states ($\widetilde{\mathcal{\varphi}}^t_D$) and share them with the E2E orchestrator after storing them. During this reduction, the domain state may be downsized from $\mathcal{S} \times (\mathcal{N} \times \mathcal{P})^2 \times \mathcal{M}_D \times \mathcal{T}_D$ to $\mathcal{S} \times \mathcal{N} \times \mathcal{L} \times \widetilde{\mathcal{M}}_D$\footnote{Let $\mathcal{N}$ denotes the number of network devices in the network domain or the number of service instances in the computing domain, $\mathcal{L}$ signifies the number of network links, and $\mathcal{M}_D$ stands for the quantity of metrics measured per resource.}, where $\widetilde{\mathcal{M}}_D$ represents the size of the domain-level service indicators set. $1$ to $\mathcal{M}_D + \widetilde{\mathcal{M}}_R$ indicators may be included in this set, affecting the size and precision of reduced states. The final task involves conveying instructions from the E2E orchestrator to underlying resources, adhering to specific standards and data models, facilitated through the configuration functionality.

\begin{figure*}[!t]
\centering
\includegraphics[width=1\textwidth]{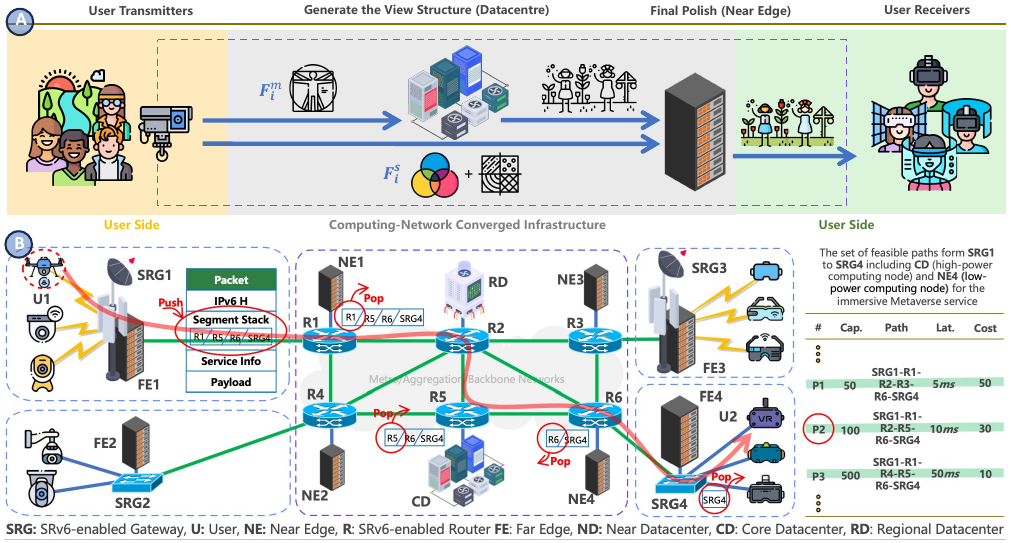}
\caption{A) A typical Metaverse use case incorporating an ACNC-enabled holographic meeting, and B) the demonstration of deterministic service provisioning enabled by SRv6 in ACNC.}
\label{fig4}
\end{figure*}

\vspace{10px}
\subsubsection{The E2E Orchestrator}\label{sss_e2e} \hfill \vspace{3px} \\  \indent 
The primary responsibility of the E2E orchestrator is to organize service instances and potential network paths for the next time slot. First, it aggregates reduced states from both domains to form a comprehensive system state, denoted as \(\varphi^t_E\), which is stored in the memory bank (depicted in Figure~\ref{fig3}). Since the network and computing orchestrators prepare the bindings of service instances with their respective resources, the E2E orchestrator concatenates these to generate a unified service-network-computing binding for all services, providing a system-wide view. Simultaneously, \(\varphi^t_E\) is labeled with a context from a set of contexts of size \(\mathcal{C}\). To achieve this, we first process the state using a GNN, then apply Vector Quantization (VQ) to the GNN's output to create embeddings for each state, grouping similar states into the same context based on the distance between the input state and the available contexts \cite{liu2024vector}. Each context corresponds to a cluster of similar states and is stored in the memory bank. Subsequently, the E2E orchestrator optimizes the reward functions of PoAs to align with system-level objectives—for example, selecting cheaper allocations when the objective is minimizing cost. When the accuracy of allocations in PoAs falls below a predetermined threshold, the Adaptive Resonance Theory (ART) is utilized to update the available set of contexts, which comprises two bidirectionally connected layers: an input layer (F1) and a recognition layer (F2) \cite{czmil2024empirical}. Upon presenting a state to F1, neurons in F2 are activated based on the similarity between the input and stored context weight vectors, potentially leading to the generation of new contexts as necessary. The implementation of ART within the framework of VQ facilitates the dynamic adjustment of the context set.

After preparing the system state and determining the current context, Long Short-Term Memory (LSTM) and Gated Recurrent Unit (GRU) models are employed to predict the next system state by analyzing historical temporal data from the last \(\mathcal{T}_E\) system states \cite{NIU20231}. These models incorporate recurrent connections and maintain an internal memory, enabling them to capture both short-term and long-term dependencies in the data for effective future state forecasting. A check is then performed on the predicted next state to detect any potential infeasibilities based on the requirements of the implemented services. Upon detecting infeasibility, the E2E orchestrator employs a RL agent to efficiently select new computing nodes for service instances. The agent uses GNN layers to process the system state represented as a graph-structured data. It is pre-trained to select a computing node for each service instance in each specific system state, optimizing for the system objective. Based on the selected computing nodes, potential E2E paths from each PoA to the chosen computing nodes are determined. The E2E orchestrator then collaborates with the network orchestrator to inform PoAs of the updated system state, context changes, as well as their action set, and with the computing orchestrator to facilitate new instance deployments. 



\begin{figure*}[!t]
\centering
\includegraphics[width=1\textwidth]{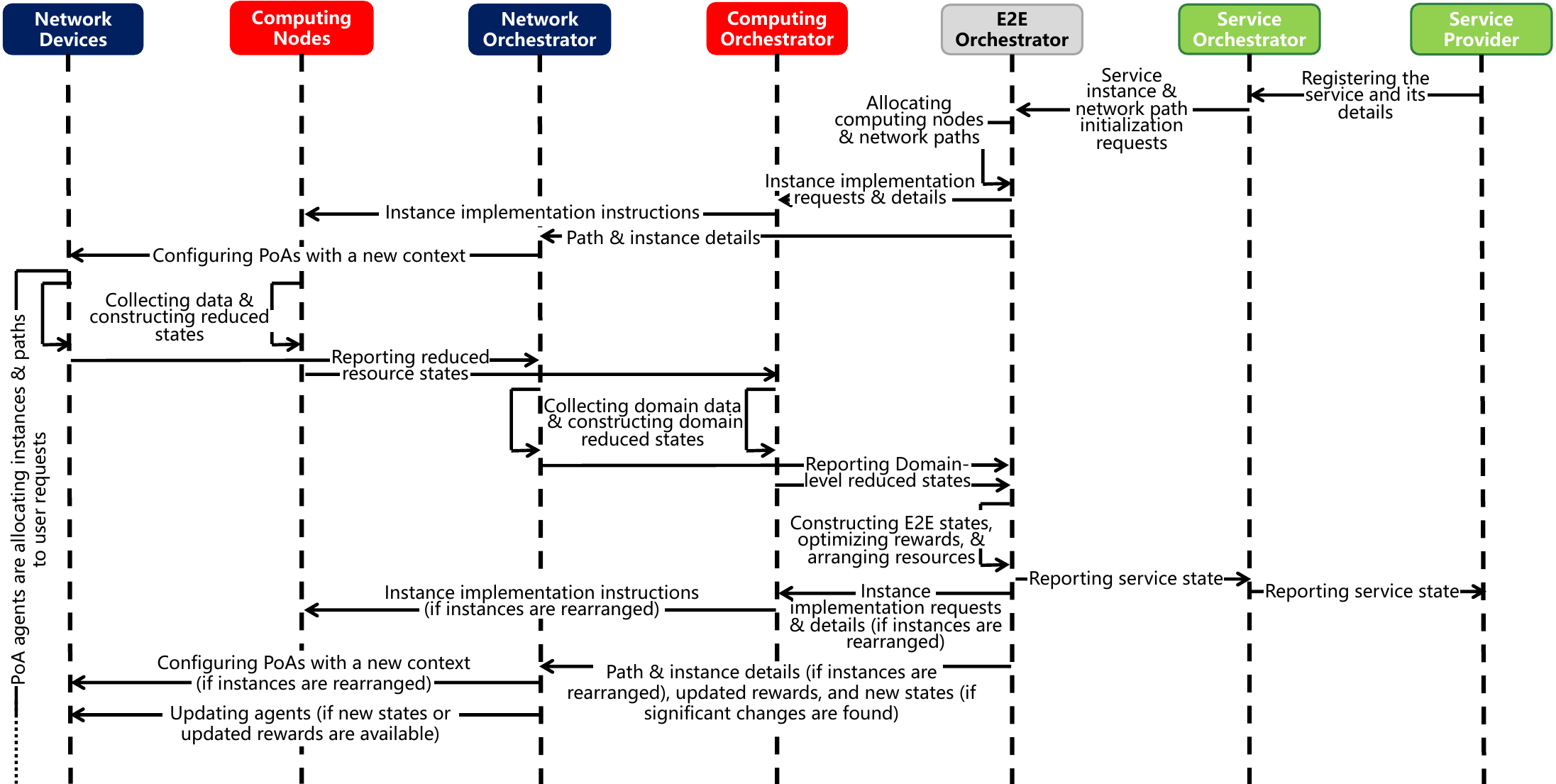}
\caption{The flowchart of E2E orchestration within the ACNC framework.}
\label{fig5}
\end{figure*}

\vspace{10px}
\subsubsection{The Service Orchestrator} \hfill \vspace{3px} \\ \indent 
This layer assumes responsibility for overseeing the service lifecycle within the ACNC framework. Its primary function involves receiving, processing, and storing requests from service providers, aiming to incorporate new services. These requests should include crucial service parameters, such as essential service instances, instance chaining maps, data models, and QoS/E requirements.
The process of deploying new service instances across the infrastructure is orchestrated by the resource arrangement component of the E2E orcherstrator, which also includes the anticipated demand reported by the service provider. Additionally, it monitors all implemented services in an E2E fashion, maintaining a real-time, abstracted representation of their quality for subsequent reporting to the respective service provider. Beyond the specifics of registered services, the memory bank houses a repository of pre-trained agents, which play a vital role in efficient decision-making, especially during the system's initial phases when there is no live data available for agent training. Furthermore, this includes simulation tools and digital twins to assess the system's efficiency and determine potential adjustments. 

\subsection{Integrating ACNC with SRv6}
To facilitate the transmission of packets for each request session, 
the involvement of network devices along the path is essential. One viable solution for conveying this information is the utilization of the Segment Routing over IPv6 (SRv6) protocol. SRv6 operates on the principles of source routing, where the routing path is encoded within the packet header as a sequenced set of segments or path hops. Each network device along the list of segments, upon packet reception, removes its associated label from the list and directs the packet toward the network device indicated by the next label. This transmission occurs directly in cases where the list is defined in a hop-by-hop manner or via the shortest path if the path is divided into segments, with one network device designated per segment. SRv6 offers the advantage of precise traffic control without the necessity for supplementary protocols or path signaling, which results in a highly scalable solution.

Consider the holographic meeting scenario. As depicted in Fig. \ref{fig4}-A, the raw data captured from each object $i$ undergoes a decomposition process into primary and auxiliary requests, specifically encompassing the transmission of positional information ($F_i^m$) and skin textures/coloring ($F_i^s$). Depending on the intended functionality, the primary requests are directed towards high-capacity computing nodes, responsible for generating a realistic view structure. Subsequently, the outcomes are forwarded to a low-capacity node, where they undergo final refinement, utilizing the traffic from auxiliary requests to reconstruct the complete scene. The final result is then delivered to the respective users. Fig. \ref{fig4}-B represents the SRv6-based routing between users U1 and U2 concerning request $F_1^m$. In accordance with the viable path options outlined in the accessible paths table, assuming that the corresponding service instances hosted on the core datacenter and NE4 offer optimal selections, three feasible paths emerge. Considering the system state, the routing agent designates P2 for $F_1^m$, allowing it to converge with the traffic from other primary requests at the core datacenter, where it proceeds to the final polishing stage on NE4. This selected path is then embedded in the request packets in the form of a segmented stack. Ultimately, the resulting output is conveyed to user U2.

\subsection{The Workflow}
The service provisioning process in ACNC, as illustrated in Fig.~\ref{fig5}, commences with the service provider registering a service and its specific characteristics within the service orchestration layer. Subsequently, the efficient computing nodes for implementing the instances of this service, along with potential E2E network paths connecting PoAs to these computing nodes, are determined by the E2E orchestrator. Instructions are then dispatched to the computing orchestrator for the implementation of these instances on the designated computing nodes, and the network orchestrator receives notifications regarding any new context or system state, if applicable. Now, predefined agents are loaded in the resource layer to manage the service requests. By conveying traffic from PoAs to computing nodes, a closed-loop process ensues, where resources, domain-level orchestrators, and the E2E orchestrator collaborate to construct the system state, leading to potential adjustments in implemented instances, network paths, and rewards assigned to PoA agents as necessary. Periodic or per-request reports are also dispatched to the service provider, presenting the QoS/E-related statistics of the implemented service.



\begin{table}[t!]
\caption{Simulation Parameters.}
\begin{center}
\begin{tabular}{|c|c|}
\hline
\textbf{Parameter} & \textbf{Value} \\
\hline
number of services & $3$ \\
number of requests per time slot & $300$ \\
number of links & $\sim \mathcal{U}\{3\mathcal{V}, 5\mathcal{V}\}$ \\
\begin{tabular}[c]{@{}c@{}}resource capacity bounds (network devices \\ and links, and computing nodes)\end{tabular} & $\sim \mathcal{U}\{250, 300\}$ mbps \\
service instance capacity bound & $20$ mbps \\
energy consumptions per capacity unit & $\sim \mathcal{U}\{10, 20\}$  \\
energy consumptions per context change & $\sim \mathcal{U}\{100, 200\}$  \\
capacity requirement per request & $\sim \mathcal{U}\{4, 8\}$ mbps\\
bandwidth requirement per request & $\sim \mathcal{U}\{2, 10\}$ mbps\\
latency requirement per request & $\sim \mathcal{U}\{1, 3\}$ ms\\
packet size per request & $1$ \\
profit per request & $\mathcal{U}\{5, 15\}$ \\
\hline
\end{tabular}
\label{T_SML}
\end{center}
\end{table}

\section{Performance Evaluation} \label{s_eva}
\subsection{Settings}
In this section, we conduct an evaluation of the proposed ACNC framework, assessing its efficacy in addressing the joint challenge of service instance placement and assignment, along with path selection in accordance with the communication pattern delineated in Section \ref{ss_arch}. The primary constraints involve the capacity and bandwidth prerequisites of requests, coupled with their admissible E2E latency threshold. The objective function is to maximize the cumulative profit, formulated as an ascending function of the number of fulfilled requests, while concurrently minimizing total energy consumption. This energy metric is construed as an ascending function of the network traffic and computing tasks managed by network devices and computing nodes, respectively. To tackle this optimization problem, we employ a Double Deep Q-Learning (DDQL) agent deployed on edge network devices, leveraging linear Deep Neural Networks (DNNs) \cite{10207694}. Furthermore, we extend this approach using Graph Neural Networks (GNNs), denoted as DDQL-GNN, to explore the impact of incorporating graph-shaped system states (generated by the E2E orchestrator) in the decision-making process. Our evaluation encompasses a comparative analysis with random allocation (RND) and optimum allocation (OPT) for a comprehensive assessment of their respective efficiencies. RND entails the random placement and assignment of instances and traffic routing, while OPT represents results obtained by solving the mixed-integer linear programming formulation of the problem using CPLEX 12.10.

\begin{figure}[!t]
\centerline{\includegraphics[width=2.3in]{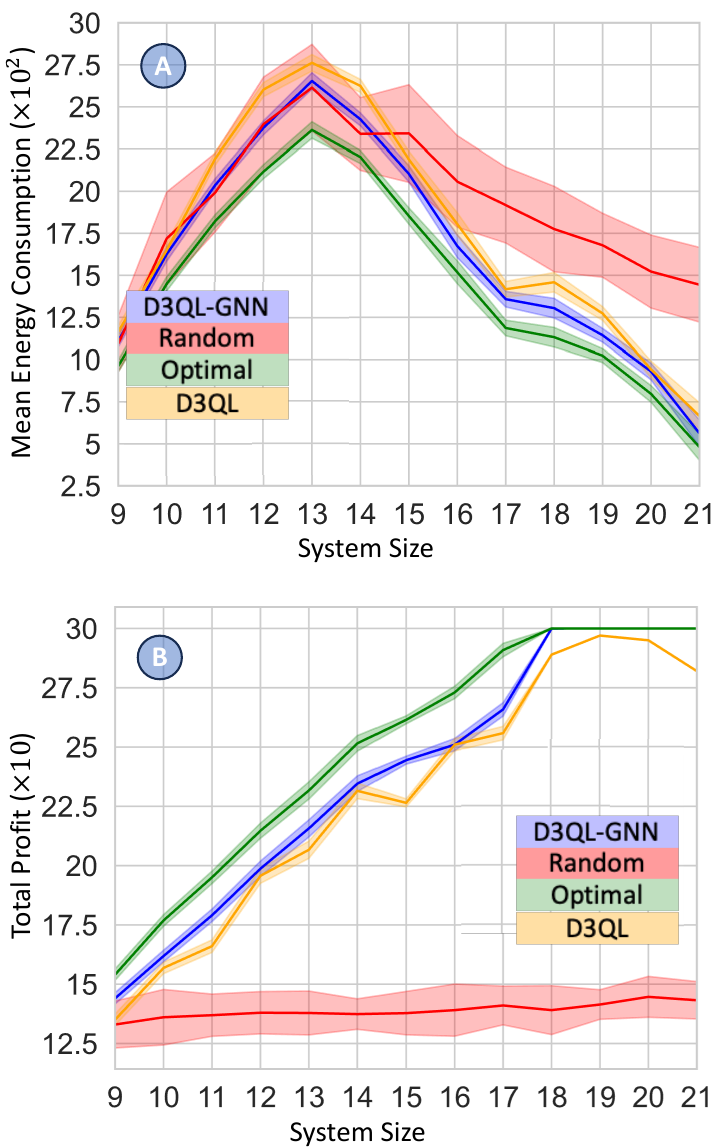}}
\caption{The mean energy consumption of supported requests (A) and the total profit (B) vs. the system size. Note that the increase in system size involves the generation of a new system graph. Specifically, resources with high energy consumption are integrated into the graph from $10$ to $13$, those with moderate energy consumption are introduced from $14$ to $17$, and the remaining resources from $18$ to $21$ exhibit low energy consumption.} \label{fig6}
\end{figure}

\subsection{Results Analysis}
The simulation results are illustrated in Fig.~\ref{fig6} for varying system sizes, denoted by $\mathcal{V}$. Each $\mathcal{V}$ represents the number of network devices, each connecting a computing node to the infrastructure. The presented results for each $\mathcal{V}$ and technique represent the mean performance over 50 time slots, where each incorporates randomly generated requests and infrastructures, adhering to uniform parameters detailed in Table \ref{T_SML}. This approach introduces dynamism into the simulations. In the figure, it's apparent that OPT serves as an upper performance bound, while RND serves as the lower bound. Notably, when all resources have high energy consumption rates or are fully occupied (with $\mathcal{N} \leq 13$), RND demonstrates a similar energy consumption pattern to DDQL-based techniques. However, RND's support rate is limited due to its lack of intelligence and feasibility checks. In contrast, DDQL-GNN excels in both scenarios, achieving near-optimal results by prioritizing high-capacity resources with minimal energy consumption, especially when multiple choices are available for each request ($\mathcal{N} \geq 13$). However, DDQL exhibits less efficiency and stability compared to DDQL-GNN, primarily due to its inferior state decoding capability.

\section{Challenges \& Future Research Directions}\label{s_chl}

\subsection{Model Pre-Training: Digital Twin Integration}

A promising direction for future research in the ACNC framework is the exploration of advanced pre-training methodologies for models and agents. Building models from scratch for each context can lead to suboptimal decision-making due to the frequent changes in the data plane and the diversity of technologies involved. Integrating digital twins emerges as a valuable strategy. By incorporating digital twins into the pre-training process and leveraging historical data, models can perform effectively in real-world situations from the outset. Key steps include identifying system components, modeling their behavior, collecting relevant data, constructing request and traffic trends, and replicating the system's behavior when all elements are integrated. A thorough investigation of these steps is essential for generating efficient pre-trained models within the ACNC framework.

\subsection{Complexities in E2E ACNC Evaluation}
In contrast to conventional communication systems, where E2E performance metrics are primarily influenced by predictable networking devices and protocols, evaluating system-wide performance in the ACNC framework poses a complex challenge. The dynamic allocation of numerous network devices and computing nodes within ACNC, coupled with the unpredictable nature of ML-based decision-making, complicates pinpointing precise sources of performance degradation. Therefore, diverse services or users may perceive performance variations differently. 
Pioneering research initiatives aiming to establish a theoretical approach for measuring E2E performance in scenarios with multiple ML agents introduced to the data path will be crucial for the comprehensive utilization of ACNC.

\subsection{ML Efficiency: Balancing Precision and Cost}
In the process of constructing system states and in all optimization tasks of the E2E orchestrator, adjustable parameters play a crucial role in balancing precision and cost. 
For instance, increasing the number of state features and indicators, as well as the history window size, provides more detailed state data. While this enhances the precision of ML agents' decisions, it necessitates increased storage for memory banks, higher network bandwidth and longer time for vertical data exchange, and additional computing capacity and time for training and decision-making. Adjusting these parameters is a prospective research avenue to maximize precision in resource allocation while minimizing network and computing costs. Dynamically navigating the trade-offs between precision and cost in ML-based decision-making represents a pivotal research initiative with profound implications for future applications.

\section{Conclusion}\label{s_con}
This paper introduced the ACNC framework, comprising service, E2E, domain, and resource building blocks. The service orchestrator registers and monitors services for ACNC, while the E2E orchestrator, the core of ACNC, provides functionalities such as E2E system state construction, context detection, reward optimization, and resource management. This orchestrator is linked to resources through domain orchestrators (network and computing), which construct short-term, long-term, and reduced domain states using received state information and configure their resources. An analysis of a Metaverse scenario demonstrated how ACNC, assisted by SRv6-enabled infrastructure, can be utilized. The workflow within ACNC was outlined, and simulations showed its superior performance compared to optimal and random strategies. The paper concluded by discussing challenges and future research directions related to digital twin integration, E2E performance evaluation, and the balance between precision and cost.

\section*{Acknowledgment}
This work is partially supported by the European Union’s HE research and innovation program HORIZON-JUSNS-2023 under the 6G-Path project (Grant No. 101139172) and the European Union’s Horizon 2020 Research and Innovation Program through the aerOS project under Grant No. 101069732. The paper reflects only the authors’ views, and the European Commission bears no responsibility for any utilization of the information contained herein.



\bibliographystyle{IEEEtran}
\bibliography{Bibliography}




\end{document}